\documentclass[10pt,conference]{IEEEtran}
\IEEEoverridecommandlockouts
\usepackage{cite}
\usepackage{amsmath,amssymb,amsfonts}
\usepackage{algorithmic}
\usepackage{graphicx}
\usepackage{textcomp}
\usepackage{xcolor}
\usepackage{xurl}
\def\BibTeX{{\rm B\kern-.05em{\sc i\kern-.025em b}\kern-.08em
    T\kern-.1667em\lower.7ex\hbox{E}\kern-.125emX}}

\usepackage{array}

\newcolumntype{L}[1]{>{\raggedright\arraybackslash}p{#1}}
\newcolumntype{C}[1]{>{\centering\arraybackslash}p{#1}}
\newcolumntype{R}[1]{>{\raggedleft\arraybackslash}p{#1}}
\newcolumntype{M}[1]{>{\centering\arraybackslash}m{#1}}

\usepackage{multirow}
\usepackage{multicol}
\usepackage[most]{tcolorbox}
\usepackage[normalem]{ulem}
\usepackage{enumitem}

\begin{document}

\title{In the Driver's Seat: A Multi-Company Study on the Reality of Autonomous Driving System Testing\\

\thanks{This work was supported in part by the Wallenberg Foundation and WASP Postdoctoral Scholarship Program - KAW 2023.0474.}
}


\author{
\IEEEauthorblockN{
Qunying Song\IEEEauthorrefmark{1},
Yuan Gao\IEEEauthorrefmark{2},
Johannes Betz\IEEEauthorrefmark{2},
Dietmar Pfahl\IEEEauthorrefmark{3},
Mohammad Reza Mousavi\IEEEauthorrefmark{4},
Federica Sarro\IEEEauthorrefmark{1}
}
\IEEEauthorblockA{
\IEEEauthorrefmark{1}University College London, London, UK\\
\IEEEauthorrefmark{2}Technical University of Munich, Munich, Germany\\
\IEEEauthorrefmark{3}University of Tartu, Tartu, Estonia\\
\IEEEauthorrefmark{4}King's College London, London, UK\\
Email: \{qunying.song, f.sarro\}@ucl.ac.uk,
\{yuan\_avs.gao, johannes.betz\}@tum.de,\\
dietmar.pfahl@ut.ee, 
mohammad.mousavi@kcl.ac.uk
}
}

\maketitle

\begin{abstract}
Autonomous driving systems (ADS) are increasingly deployed in complex real-world environments, creating strong demands for rigorous testing to establish their functionality and safety. However, ADS testing remains difficult because practitioners must determine not only what and how to test, but also whether the resulting evidence is sufficient to support progression and release decisions. To understand how this is handled in practice, we conducted semi-structured interviews with experts involved in ADS development and testing across nine companies in six countries. Through thematic analysis, we examined how practitioners organize ADS testing to produce and assess evidence, and what limits confidence in that evidence. Our findings show that industrial ADS testing can be understood as an evidence-building process involving three recurring activities: planning what evidence is needed, producing and evaluating evidence across complementary testing environments, and deciding when testing is sufficient. We further identified recurring limitations concerning scenario adequacy, simulation and deployment fidelity, test execution and observability, and assurance and decision sufficiency. Practitioners reported several approaches for addressing these limitations, although important gaps remain. Based on these findings, we propose an evidence-centered closed-loop testing framework that connects testing intent, scenario selection, evidence production, acceptance decisions, and operational feedback. The study provides an industry-grounded account of how ADS testing is organized, where confidence in current testing breaks down, and how testing processes can be structured more systematically.
\end{abstract}


\section{Introduction}
\label{sec:intro}

Autonomous driving systems (ADS) have been rapidly advancing and deployed in commercial services across different geographical regions~\cite{liao2025advancing}. These advances require testing practices to co-evolve in order to effectively validate system functionality and safety~\cite{sun2021scenario}. However, ADS testing remains non-trivial, involving diverse approaches, environments, tools, and evaluation criteria~\cite{tang2023survey}. Unlike conventional software testing, ADS testing must account for diverse driving situations, interactions with other road users, and uncertainties in system behavior. Practitioners must therefore determine not only what and how to test, but also what evidence should be produced and when that evidence is sufficient to support progression or release decisions.

Existing research has examined individual aspects of ADS testing, including testing techniques, tools, scenarios, and safety assurance~\cite{tang2023survey, liao2025advancing, khan2023safety}. However, there remains limited empirical understanding of how these elements are organized and combined in industrial practice to build confidence in ADS safety and performance~\cite{song2026research}. In particular, although previous studies have identified challenges such as scenario coverage, simulation fidelity, and testing complexity~\cite{kang2019test, beringhoff2022thirty}, less is known about how such limitations affect confidence in testing evidence and how practitioners address them under real industrial constraints. As a result, it remains unclear not only which practices are used, but also how they work together to support the safe release and deployment of ADS.

To address these gaps, we conducted an interview study with industry practitioners directly involved in ADS development and testing. Rather than describing individual practices and challenges, we examine ADS testing as an evidence-building and decision-making process. Specifically, we investigate the following research questions:

\begin{enumerate}[label=\textbf{RQ\arabic*:},
                  leftmargin=*,
                  topsep=6pt,
                  itemsep=4pt,
                  parsep=0pt,
                  partopsep=0pt]
    \item How do practitioners organize ADS testing to produce and assess evidence for system safety and performance?
    \item What limits confidence in the evidence produced by current ADS testing, and how do practitioners address these limitations?
\end{enumerate}

In this paper, we use \textit{testing evidence} broadly to refer to observed system behaviors, logs, measurements, coverage, and other test outcomes used to support safety assessment, testing progression, release, and deployment decisions.

We conducted semi-structured interviews~\cite{rowley2012conducting, runeson2009guidelines} with nine experts from nine companies across six countries, covering diverse roles and experience in ADS development and testing. We analyzed the interview data using thematic analysis~\cite{cruzes2011recommended, defranco2017content} and synthesized the findings into a thematic model. The interviews were intentionally open-ended to allow participants to describe testing practices, problems, and experiences from their own industrial contexts, and participants were given the opportunity to review and revise their transcripts.

Our findings show that industrial ADS testing can be understood as a recurring process of planning what evidence is needed, producing and evaluating evidence across complementary testing approaches, activities, and tools, and deciding when the accumulated evidence is sufficient. This process is primarily supported by scenario-based and X-in-the-loop testing, but confidence in the resulting evidence is limited by persisting issues related to scenario adequacy, simulation and deployment fidelity, test execution and observability, and assurance and decision sufficiency. Practitioners reported several approaches for addressing these limitations, although important gaps remain, particularly in determining whether testing evidence is sufficiently comprehensive and convincing. Based on these findings, this study makes three main contributions:

\begin{itemize}
    \item \textbf{Empirical insights.} We provide an industry-grounded characterization of ADS testing as an \textit{evidence-building process}, showing how practitioners plan testing, produce and evaluate evidence across testing environments, and make progression and acceptance decisions.
    \item \textbf{Challenges and responses.} We synthesize the \textit{major limitations} that undermine confidence in ADS testing evidence, together with practitioner reported approaches for addressing them and remaining unresolved gaps.
    \item \textbf{Actionable framework.} We derive an \textit{evidence-centered closed-loop testing framework} that connects testing intent, scenario selection, evidence production, acceptance decisions, and operational feedback, providing actionable guidance for structuring ADS testing and identifying where additional evidence is needed.
\end{itemize}

\section{Related Work}

\subsection{Reviews of ADS Testing Research}

A substantial body of work has reviewed ADS testing research from different perspectives. Zhang et al.~\cite{zhang2022finding} and Ding et al.~\cite{ding2023survey} focused on critical scenario identification, while Riedmaier et al.~\cite{riedmaier2020survey} examined scenario-based safety assessment. Tang et al.~\cite{tang2023survey} provided a broader review of ADS testing techniques, challenges, and future trends. Other studies have surveyed testing tools, datasets, simulators, and platforms~\cite{ji2021perspective, kang2019test, ma2021traffic, rosique2019systematic, cai2022survey}, and more recent work has examined the emerging role of generative AI and foundation models in ADS testing~\cite{song2025generative, zhao2026survey, tian2025large, gao2026foundation, wu2026foundation}. These studies draw primarily on academic literature or publicly available information and provide valuable syntheses of existing techniques and resources.

However, such reviews provide limited insight into how these individual techniques and resources are actually organized and combined in industrial testing processes. Our study complements these literature by directly examining practitioners' experiences and focusing on how testing is planned, how evidence is produced and evaluated across different testing environments, and how practitioners determine whether that evidence is sufficient for progression and acceptance decisions.

\subsection{Empirical Studies of Industrial ADS Testing}

Several empirical studies have investigated ADS testing in industry. Beringhoff et al.~\cite{beringhoff2022thirty} interviewed practitioners and identified 31 challenges in ADS testing. Song et al.~\cite{song2024empirically, song2024industry} examined industrial practices and challenges specifically related to scenario-based testing. Lou et al.~\cite{lou2022testing} combined interviews, questionnaires, and a systematic literature review to investigate testing practices and gaps between academic research and industrial needs. Similarly, Liao et al.~\cite{liao2025advancing} combined survey and literature review to study industrial demands, challenges, and future directions in ADS testing.

Although these studies provide important empirical insights, they largely characterize particular practices, challenges, or research--industry gaps as separate findings. In contrast, our study focuses on how these elements fit together within the broader testing process. We characterize industrial ADS testing as an evidence building process in which practitioners determine what evidence is needed, produce and evaluate it across complementary testing approaches, activities, metrics, and tools, and decide when the accumulated evidence is sufficient. We further examine the limitations that reduce confidence in this evidence and the approaches practitioners use to address them. Building on these empirical findings, we synthesize an evidence-centered closed-loop testing framework that connects testing intent, scenario selection, evidence production, acceptance decisions, and operational feedback.

Overall, our study differs from prior work not simply by covering a broader set of contemporary practices or challenges, but by providing an integrated view of \textit{how industrial ADS testing builds evidence, where confidence in that evidence breaks down, and how these these findings and future directions can be translated into a structured testing process}.

\section{Research Method}
\label{sec:method}

\begin{figure*}[tbp]
    \centering
    \includegraphics[width=\textwidth]{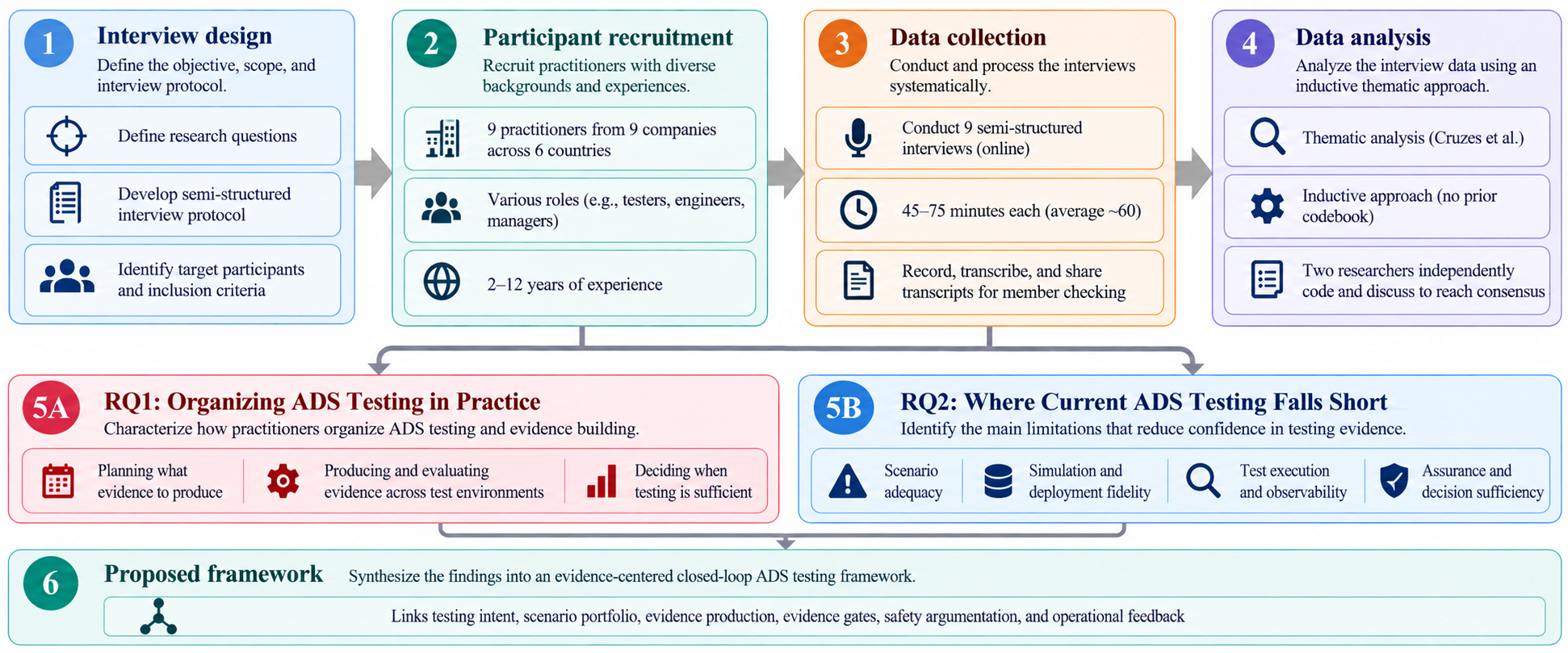}
    \caption{Overview of the interview study, from participant recruitment and data collection to thematic analysis, the two research questions, and synthesis of the evidence-centered closed-loop testing framework.}
    \label{fig:methodology}
\end{figure*}

As summarized in Figure~\ref{fig:methodology}, our study follows four main steps: participant recruitment, semi-structured interviews, inductive thematic analysis, and synthesis of the findings around the two research questions and the proposed framework.

\begin{table*}[tbp]
\centering
\caption{Overview of interview participants (P1--P9) and their companies (C1--C9). Experience includes total industry experience, with experience in autonomous driving shown in parentheses. Company information is based on LinkedIn profiles.}
\begin{tabular}{|C{0.035\textwidth}|C{0.16\textwidth}|C{0.11\textwidth}|C{0.09\textwidth}|C{0.055\textwidth}|C{0.28\textwidth}|C{0.09\textwidth}|}
    \hline
    \textbf{\#} &
    \textbf{Role} &
    \textbf{Experience} &
    \textbf{Location} &
    \textbf{Company} &
    \textbf{Industry} &
    \textbf{Size} \\
    \hline
    P1 & System Analyst & 9 (9) years & Sweden & C1 & Motor Vehicle Manufacturing & 10,001+ \\
    \hline
    P2 & Chief Engineer & 13 (12) years & China & C2 & Motor Vehicle Manufacturing & 10,001+ \\
    \hline
    P3 & System Tester & 14 (4) years & China & C3 & Motor Vehicle Manufacturing & 10,001+ \\
    \hline
    P4 & Researcher & 10 (10) years & Sweden & C4 & Motor Vehicle Manufacturing & 10,001+ \\
    \hline
    P5 & System Tester & 2 (2) years & Germany & C5 & Software Development & 201--500 \\
    \hline
    P6 & Product Engineer & 4 (2) years & China & C6 & Motor Vehicle Manufacturing & 10,001+ \\
    \hline
    P7 & Senior Engineer & 6 (3) years & UK & C7 & Technology, Information and Internet & 1,001--5,000 \\
    \hline
    P8 & Managing Director & 10 (7) years & Japan & C8 & Automotive & 2--10 \\
    \hline
    P9 & Engineering Manager & 10 (10) years & Belgium & C9 & Automation Machinery Manufacturing & 10,001+ \\
    \hline
\end{tabular}
\label{tab:participants}
\end{table*}

\subsection{Participant Recruitment}

We used multiple sampling approaches to recruit interview participants. \textit{Convenience sampling}~\cite{baltes2022sampling, ghazi2018survey} was the most effective, through which we invited practitioners known to the research team and contacts from previous industry collaborations, recruiting seven participants. To broaden the sample, we used \textit{purposive sampling}~\cite{baltes2022sampling, ghazi2018survey} by approaching relevant companies and inviting their testing experts, which yielded two additional participants. In total, nine practitioners were interviewed. We also attempted self-selection sampling through social-media calls (on LinkedIn and X) and snowball sampling by asking interviewees to recommend additional candidates, but neither yielded further participants.

\subsection{Interviewee Characteristics}

As shown in Table~\ref{tab:participants}, the nine interviewees came from nine companies involved in autonomous driving across six countries. Most companies were large enterprises, except for one medium-sized company (C5) and one small company (C8). Participants held diverse roles, including testers, engineers, and managers. Four participants (P3, P5--P7) had 2--4 years of experience in ADS development and testing, while five (P1--P2, P4, P8--P9) had 7--12 years. Overall, the sample covered companies of different sizes and geographic locations and participants with diverse roles and experience levels.

\subsection{Interview Protocol}

\begin{table}[tbp]
\centering
\caption{Condensed interview guide. The full interview guide, including detailed explanations and examples, is available on Zenodo~\cite{song_2026_21408632}.}
\begin{tabular}{L{0.95\columnwidth}}
    \hline
    \textbf{Part I -- Background} \\
    (1) What is your role and experience in ADS testing? \\
    \hline
    \textbf{Part II -- Testing Practices} \\
    (2) Which ADS have you worked on, and what are their key characteristics? \\
    (3) What testing processes and activities are commonly used? \\
    (4) What are the goals and acceptance criteria for each activity? \\
    (5) What approaches, tools, and metrics are typically used? \\
    \hline
    \textbf{Part III -- Challenges and Potential Solutions} \\
    (6) What are the main challenges in ADS testing? \\
    (7) Which challenge is the most critical, and why? \\
    (8) What solutions or future directions are worth exploring? \\
    \hline
    \textbf{Part IV -- Future Outlook} \\
    (9) How do you see ADS testing evolving in the future? \\
    \hline
\end{tabular}
\label{tab:questions}
\end{table}

Before each interview, we provided participants with a brief description of the study, its goals and scope, the interview tool and expected duration, and a consent form covering voluntary participation, the right to skip questions or withdraw at any time, and data anonymization. Interviews were conducted via Microsoft Teams and lasted 45--75 minutes, averaging approximately 60 minutes. At the beginning of each interview, we reiterated the consent information and obtained permission to record the session. We then used the nine questions in Table~\ref{tab:questions} to guide a semi-structured discussion~\cite{runeson2009guidelines, rowley2012conducting}, with follow-up questions asked where appropriate. The questions covered participant background, ADS testing practices, testing challenges, and future outlooks.

\subsection{Qualitative Data Analysis}

After each interview, we saved the recording and transcript and checked the transcript against the recording to correct missing or unclear content. Participants were also invited to review and revise their transcripts. We then followed the thematic analysis guidelines of Cruzes et al.~\cite{cruzes2011recommended} using an \textit{inductive} approach without a predefined codebook. Relevant segments were coded, and codes across interviews were iteratively merged and grouped into themes and higher-level themes to form a coherent thematic model covering testing practices, challenges, potential solutions, and future outlooks. 

The original themes remained the empirical basis of the analysis. For this paper, we further synthesized them around the two research questions rather than reporting every theme independently, as summarized in Figure~\ref{fig:methodology}. Practice related themes were organized according to their role in planning evidence, producing and evaluating evidence, and deciding testing sufficiency. Challenges and reported solutions were grouped according to the aspect of testing evidence they affect, while future outlook themes are discussed in Section~\ref{sec:discussion:future_trends}. Participant identifiers are retained selectively when a concrete example or distinct viewpoint is important, while recurring observations are synthesized across participants.

\section{RQ1: Organizing ADS Testing in Practice}
\label{sec:rq1}

\begin{figure*}[tbp]
    \centering
    \includegraphics[width=\textwidth]{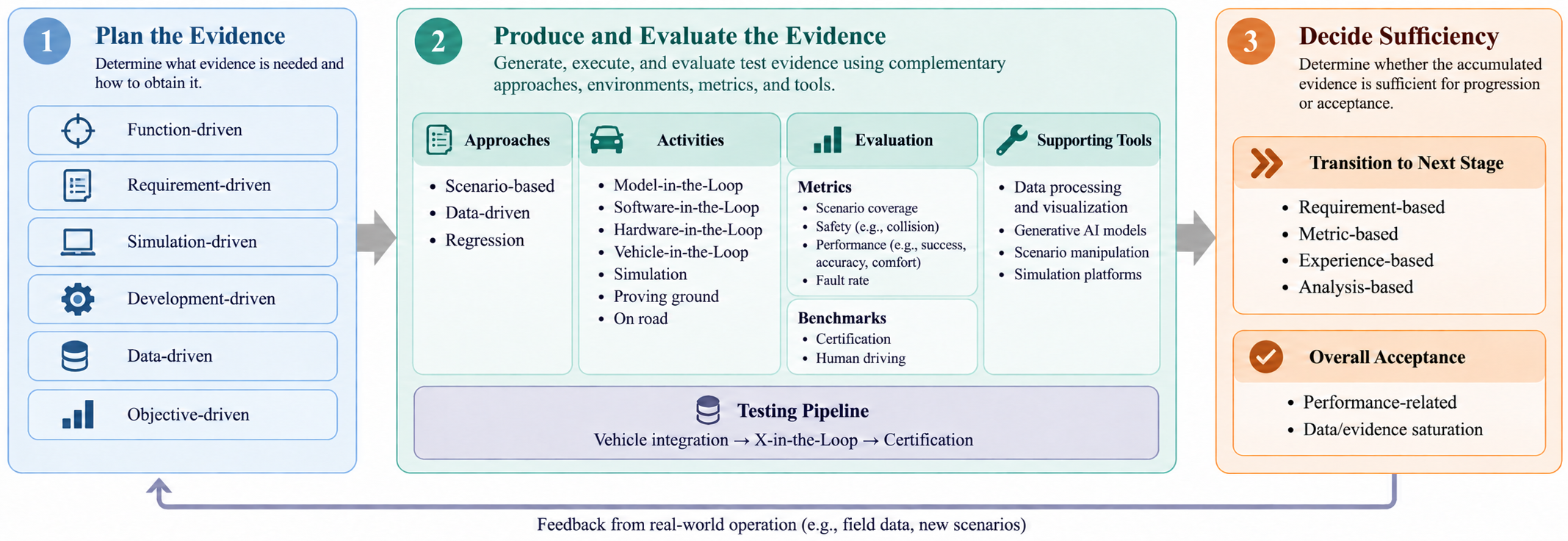}
    \caption{Organization of industrial ADS testing synthesized from the interview findings: planning the required evidence, producing and evaluating evidence across complementary testing activities, and deciding when testing is sufficient.}
    \label{fig:practices}
\end{figure*}

Our interviews suggest that industrial ADS testing can be understood as an evidence-building process with three recurring activities: determining what evidence is needed, producing and evaluating that evidence across complementary testing environments, and deciding whether the accumulated evidence is sufficient to progress or accept the system, as summarized in Figure~\ref{fig:practices}.

\subsection{Planning What Evidence to Produce}

Participants described ADS testing plans as highly dependent on the system and testing context rather than following a fixed strategy. Important considerations included the function under test, testing requirements, operational design domain (ODD), collected driving data, development maturity, and the capabilities and limitations of available test environments.

Several participants emphasized that the \textit{function and requirements} largely determine what should be tested. P5 and P6 explained that different functions, such as parking and highway driving, require different test scenarios, environments, and evaluation criteria. P4 similarly described planning testing from functional and system requirements, while considering whether a particular test environment can provide the required evidence. For example, software simulation may be sufficient for some functional behaviors but cannot expose hardware failures, delays, or other physical effects that require Hardware-in-the-Loop (HiL) or Vehicle-in-the-Loop (ViL) testing.

Testing is also becoming increasingly \textit{data-driven}. P2 observed that, while traditional rule-based systems can often be represented through explicit object-level models, modern learning-based ADS increasingly depend on large-scale multimodal sensor data. 
The \textit{development maturity} of the system also shapes the testing plan. P3 described testing as an iterative process in which additional functions are gradually introduced during development and across release stages, evaluated, and refined as the vehicle and its interconnected systems mature.

Two further considerations cut across these strategies. P1 described an \textit{objective-driven} approach in which testing is planned around the objectives (e.g., safety, quality) to be demonstrated and the evidence required to support it. P4 and P9 also emphasized moving suitable testing activities toward \textit{simulation} because physical testing is costly, time-consuming, and potentially hazardous. However, this shift is selective rather than absolute: physical testing remains necessary when hardware behavior, communication delays, sensor effects, or environmental fidelity are central to the evidence being sought. Thus, practitioners select environments according to both their efficiency and ability to support the intended testing objective.

Overall, these accounts indicate that test planning is not simply the selection of a predefined sequence of activities or follow a fixed testing pipeline. Rather, practitioners determine what evidence is required based on the function, requirements, operating context, available data, and system maturity, and then select testing approaches and environments accordingly.

\begin{tcolorbox}[
    enhanced,
    colback=black!5,
    frame hidden,
    borderline west={3pt}{0pt}{black},
    sharp corners,
    boxsep=0pt,
    left=10pt,
    right=10pt,
    top=8pt,
    bottom=8pt
]
\textbf{\textit{Key finding 1:}} ADS testing plans are context-dependent rather than fixed, with the required evidence shaped by the function under test, requirements, operating context, available data, system maturity, and testing capabilities.
\end{tcolorbox}

\subsection{Producing and Evaluating Evidence Across Test Environments}

Practitioners primarily described \textit{scenario-based testing} as a widely used approach for systematically evaluating ADS behavior under various driving situations. Scenarios may originate from standards and regulations, expert knowledge, real-world data, or synthetically generated scenarios, and are selected according to the function and testing objective. Participants also described increasingly \textit{data-driven testing} approaches, in which real-world logs are replayed, reconstructed, or used to extend relevant test scenarios.

Evidence is then produced across complementary testing environments and activities. Nearly all participants (P1--P5, P7--P9) described \textit{X-in-the-Loop} testing spanning model-, software-, hardware-, and vehicle-level evaluation. Earlier activities provide scalability and easier isolation of possible faults, whereas later activities progressively introduce real software, hardware interactions, and physical operating conditions. \textit{Simulation} is therefore widely used for large-scale scenario execution, and rare or hazardous situations. 
\textit{Proving ground} testing adds real sensors, vehicles, and controlled environmental conditions, and \textit{public road} testing exposes the integrated system to situations and interactions that may not be sufficiently represented in simulation. The activities are not always strictly sequential; practitioners may return to earlier environments as software and scenarios evolve. Different environments therefore produce different forms of evidence rather than substituting for one another. 

Participants further described supporting tools for data \textit{processing}, \textit{synchronization}, \textit{visualization}, \textit{scenario manipulation}, and \textit{simulation}, which enable the broader testing process. The resulting evidence is evaluated using diverse measures such as \textit{collision occurrence}, \textit{scenario coverage}, \textit{success and fault rates}, \textit{accuracy}, \textit{interventions}, and \textit{comfort}. P8 further distinguished between formal benchmarks, such as predefined \textit{certification} testing and scenarios, and broader reference baselines, such as \textit{human driving} performance.

Evidence production also depends on activities surrounding the execution of test scenarios. P5, working for a testing service provider, described \textit{vehicle integration} as a substantial prerequisite to full-vehicle testing. Vehicles may need to be equipped and configured with cameras, signal-capturing devices, GPS modules, communication components, and specialized equipment for accessing internal vehicle signals. This preparation can require considerable engineering effort before testing begins. P5 and P6 also described \textit{certification testing} as an external evaluation activity. Failures during certification may require tests to be repeated or rescheduled, making rapid diagnosis and corrective action important before another attempt.

Taken together, the interview findings show that industrial ADS testing relies on combining evidence from multiple environments and activities rather than following a universally fixed sequence or standardized process. Virtual simulation environments provide efficiency, scalability, and controllability, whereas hardware and vehicle-based testing expose integration effects and increasingly realistic operating conditions. Vehicle preparation, supporting tools, evaluation metrics, and certification further shape whether the resulting evidence is usable for system acceptance, release, and deployment decisions.

\begin{tcolorbox}[
    enhanced,
    colback=black!5,
    frame hidden,
    borderline west={3pt}{0pt}{black},
    sharp corners,
    boxsep=0pt,
    left=10pt,
    right=10pt,
    top=8pt,
    bottom=8pt
]
\textbf{\textit{Key finding 2:}} ADS testing combines complementary virtual and physical environments, where scalable and controllable testing is progressively complemented by evidence of hardware integration, system-level behavior, and real-world performance.
\end{tcolorbox}

\subsection{Deciding When Testing Is Sufficient}

Producing evidence alone does not determine whether testing is sufficient. Participants described different criteria for deciding whether a test activity or testing phase had produced sufficient evidence to proceed.

At the level of individual activities or stages, progression was guided by \textit{requirements, engineering experience, and quantitative measures}. P1 and P7 described using performance and coverage \textit{measures} to determine whether additional testing was needed, while P9 emphasized evaluating results against \textit{requirements} established before a testing campaign. Participants also noted that progression is not governed entirely by quantitative measurements. P1 and P2 described relying partly on accumulated \textit{engineering experience} when boundaries between sufficient and insufficient testing are unclear. Practical constraints such as time and resources can also influence how much testing is performed (P3, P6).

At the system level, acceptance depends on whether the accumulated evidence demonstrates adequate functional and safety \textit{performance}. Participants referred to criteria such as collision avoidance, success and accuracy thresholds, scenario coverage, regulatory requirements, and certification results. P1 and P7 also described a \textit{data saturation} criterion, where progression or release depends on whether sufficient driving data have been collected and whether observed behavior remains consistent with expectations over the accumulated mileage. P7 described staggered release phases with mileage targets, while P1 emphasized the difficulty of determining how much road testing data are enough to conclude that the system is mature. P8 distinguished between predefined certification scenarios, which establish a minimum performance level, and comparisons with human-driving statistics as a higher-level reference for deployed ADS performance. P1 further described structuring evidence through assurance cases such as goal structuring notation (GSN), while noting that the top-level question of what constitutes ``safe enough'' remains difficult to resolve. Thus, meeting individual performance, mileage, or certification thresholds does not necessarily establish overall testing sufficiency.

Thus, sufficiency decisions combine requirements, quantitative measures, accumulated driving data, benchmarks, safety considerations, and engineering judgment. This makes progression and acceptance an integral part of the testing process, but also exposes an important limitation: the evidence required to justify a decision is often context-dependent and cannot always be reduced to a universal threshold.

\begin{tcolorbox}[
    enhanced,
    colback=black!5,
    frame hidden,
    borderline west={3pt}{0pt}{black},
    sharp corners,
    boxsep=0pt,
    left=10pt,
    right=10pt,
    top=8pt,
    bottom=8pt
]
\textbf{\textit{Key finding 3:}} Testing sufficiency is determined through a combination of requirements, safety and performance measures, benchmarks, and engineering judgment, rather than by a single universal progression or acceptance criterion.
\end{tcolorbox}

\section{RQ2: Where Current ADS Testing Falls Short}
\label{sec:rq2}

\begin{figure*}[tbp]
    \centering
    \includegraphics[width=\textwidth]{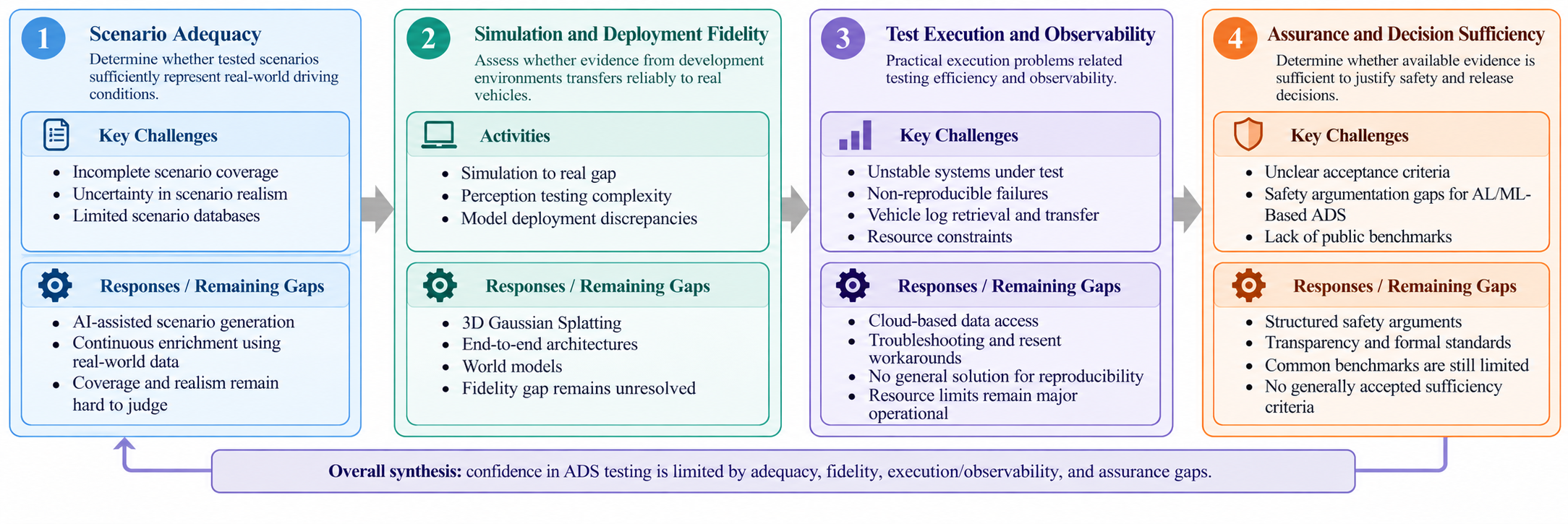}
    \caption{Four groups of challenges that limit confidence in ADS testing evidence, together with practitioner-reported responses and remaining gaps.}
    \label{fig:challenges}
\end{figure*}

As summarized in Figure~\ref{fig:challenges}, we identified 13 challenges that limit the effectiveness of ADS testing or confidence in the evidence it produces. Rather than treating them as isolated problems, we group them into four broader areas: scenario adequacy, simulation and deployment fidelity, test execution and observability, and assurance and decision sufficiency. Participants reported potential responses to several challenges, although many remain only partially addressed.

\subsection{Scenario Adequacy}

A recurring concern was whether the tested scenarios adequately represent the relevant driving space. Participants highlighted \textit{incomplete scenario coverage}, noting that rare, unexpected, or previously unseen situations are difficult to anticipate and that it remains unclear when sufficient coverage has been achieved. This is particularly problematic for learning-based ADS, whose behavior may vary in conditions not represented during development or testing.

Participants also questioned the \textit{realism of generated scenarios}. P9 noted that optimization-based or AI-generated critical scenarios can become unrealistic or unrepresentative of actual driving, while P2 highlighted the difficulty of assessing realism for complex sensor-level representations such as images and point clouds. Closely related, participants reported that existing \textit{scenario databases} may not capture the diversity and complexity encountered in real road traffic.

Practitioners described several ways of extending scenario coverage. P5 suggested combining AI-assisted scenario generation with large-scale real-world data to discover new combinations and unexpected situations. P3 similarly described continuously enriching scenario collections using road data and generated scenarios, while organizing them according to environmental and driving conditions. However, participants did not report an established method for determining when scenario coverage or realism is sufficient.

\begin{tcolorbox}[
    enhanced,
    colback=black!5,
    frame hidden,
    borderline west={3pt}{0pt}{black},
    sharp corners,
    boxsep=0pt,
    left=10pt,
    right=10pt,
    top=8pt,
    bottom=8pt
]
\textbf{\textit{Key finding 4:}} Scenario collections can be continuously expanded using real-world data and generation techniques, but practitioners still lack reliable ways to determine whether the tested scenarios are sufficiently comprehensive and realistic.
\end{tcolorbox}

\subsection{Simulation and Deployment Fidelity}

A second group of challenges concerns whether evidence produced during development accurately reflects deployed-system behavior. Participants repeatedly highlighted the \textit{simulation-to-real gap}. Differences in sensor behavior, materials, reflections, environmental conditions, and physical interactions can make simulated scenarios diverge from reality. Participants also noted that simplified representations may omit objects or effects that become important in real operation.

Participants therefore viewed \textit{simulation fidelity} as purpose-dependent rather than absolute. For example, a simulation environment that is adequate for testing planning or control behavior may still be unsuitable for evaluating perception, sensor effects, hardware timing, or other phenomena whose realism is central to the testing objective. Selecting a simulator is consequently also a question of whether its abstraction and fidelity are appropriate for the evidence being produced.

\textit{Perception testing} was considered particularly difficult because perception is influenced by many environmental factors and errors can propagate to downstream planning and control. P1 and P7 described conditions such as lighting, shadows, sensor behavior, and other physical effects that are costly or difficult to reproduce faithfully in simulation.

Participants further reported \textit{discrepancies between development and deployment environments}. P7 noted that models developed using frameworks such as PyTorch may behave differently after conversion, optimization, and integration into resource-constrained vehicle platforms, where latency, model size, and inference behavior become important.

Potential responses mainly aim to narrow these gaps rather than to eliminate them. Participants discussed more realistic simulation techniques, including 3D Gaussian Splatting, while noting remaining difficulties in reproducing physical materials, reflections, and deformation. End-to-end approaches and world models were also discussed as ways of representing richer interactions between the environment, sensors, and ADS. For deployment discrepancies, participants described dedicated integration work and techniques such as quantization, pruning, and distillation, followed by renewed simulation testing. Nevertheless, fidelity remains an unresolved limitation across the testing chain.

\begin{tcolorbox}[
    enhanced,
    colback=black!5,
    frame hidden,
    borderline west={3pt}{0pt}{black},
    sharp corners,
    boxsep=0pt,
    left=10pt,
    right=10pt,
    top=8pt,
    bottom=8pt
]
\textbf{\textit{Key finding 5:}} Evidence produced in simulation and development environments does not always transfer reliably to deployed vehicles. Emerging simulation and model integration techniques can reduce these discrepancies, but cannot yet remove the underlying fidelity gap.
\end{tcolorbox}

\subsection{Test Execution and Observability}

Even when suitable scenarios and environments are available, practical difficulties can limit the evidence that testing produces. Participants reported \textit{unstable systems under test}, where software, hardware, or integration faults interrupt testing and require repeated restarts or troubleshooting. Such failures may be intermittent and difficult to isolate, reducing testing efficiency.

Relatedly, \textit{non-reproducible failures} create substantial debugging effort. P3 described critical failures that occur only occasionally and are difficult to reproduce under identical conditions, often requiring coordination among testing teams, system owners, and suppliers. Participants did not identify an effective general solution to this problem.

\textit{Data retrieval and transfer} can further limit diagnosis. P6 described difficulties obtaining complete vehicle logs, particularly when data must be transferred manually or accessed remotely from vehicles operating in other regions. Cloud-based upload and remote access were viewed as useful for accelerating analysis and model updates, although cross-border data restrictions may require local data centers and data desensitization.

Finally, \textit{resource constraints} affect how much testing can be performed. P3 and P6 highlighted limitations in time, equipment, and development schedules, which can restrict coverage or force teams to accept known limitations. As this is largely a budgeting and planning issue, participants did not propose a specific technical solution beyond recognizing its influence on testing decisions.

\begin{tcolorbox}[
    enhanced,
    colback=black!5,
    frame hidden,
    borderline west={3pt}{0pt}{black},
    sharp corners,
    boxsep=0pt,
    left=10pt,
    right=10pt,
    top=8pt,
    bottom=8pt
]
\textbf{\textit{Key finding 6:}} Testing effectiveness is constrained not only by testing techniques, but also by system instability, failure reproducibility, data accessibility, and available resources, several of which remain primarily operational rather than technical problems.
\end{tcolorbox}

\subsection{Assurance and Decision Sufficiency}

The final group of challenges concerns whether the accumulated testing evidence is sufficient to justify safety and release decisions for the system under test. Participants highlighted \textit{unclear acceptance criteria}, particularly the difficulty of defining when an ADS is ``safe enough.'' P1 described using structured safety arguments such as GSN, but noted that determining acceptable top-level safety remains difficult and may involve ethical, legal, and organizational judgment.

Participants also identified \textit{gaps in safety argumentation} for learning-based ADS. P2 and P4 noted that established safety approaches are less straightforward for complex AI-based systems, particularly when justifying scenario selection, environmental coverage, simulation fidelity, and the behavior of less interpretable models. No generally effective solution was reported for acceptance criteria or safety argumentation.

A related problem is the \textit{lack of public and comparable benchmarks}. P8 observed that, unlike some other AI domains, ADS developers rarely share comparable performance results, sensor configurations, or evaluation procedures, making cross-company comparison difficult. Participants suggested that greater transparency, common standards, and industry-supported benchmarks could improve comparability without requiring companies to disclose proprietary systems.

\begin{tcolorbox}[
    enhanced,
    colback=black!5,
    frame hidden,
    borderline west={3pt}{0pt}{black},
    sharp corners,
    boxsep=0pt,
    left=10pt,
    right=10pt,
    top=8pt,
    bottom=8pt
]
\textbf{\textit{Key finding 7:}} The most fundamental assurance challenge is not simply producing more testing evidence, but determining whether that evidence is sufficient to support a convincing safety or release decision; common acceptance criteria and comparable benchmarks remain limited.
\end{tcolorbox}

Across the four areas, the maturity of reported responses was uneven. Participants described concrete approaches for expanding scenario collections, improving simulation realism, adapting models for deployment, and improving data access. In contrast, they reported no generally effective solutions for several problems related to failure reproducibility, top-level acceptance criteria, and safety argumentation. This distinction suggests that producing more tests or more data alone does not resolve all testing limitations; some of the remaining difficulties concern how the resulting evidence can be interpreted and justified.

The four areas are also interdependent. Scenario adequacy determines whether relevant behavior is validated; simulation and deployment fidelity affect whether resulting observations transfer to the deployed system; execution and observability determine whether failures can be captured and diagnosed; and assurance criteria determine whether the accumulated evidence is convincing enough to support a decision. Weakness in any one of these areas can therefore reduce confidence in evidence produced elsewhere in the testing process.

\section{Evidence-Centered Closed-Loop ADS Testing Framework}
\label{sec:framework}

\begin{figure*}[tbp]
    \centering
    \includegraphics[width=\textwidth]{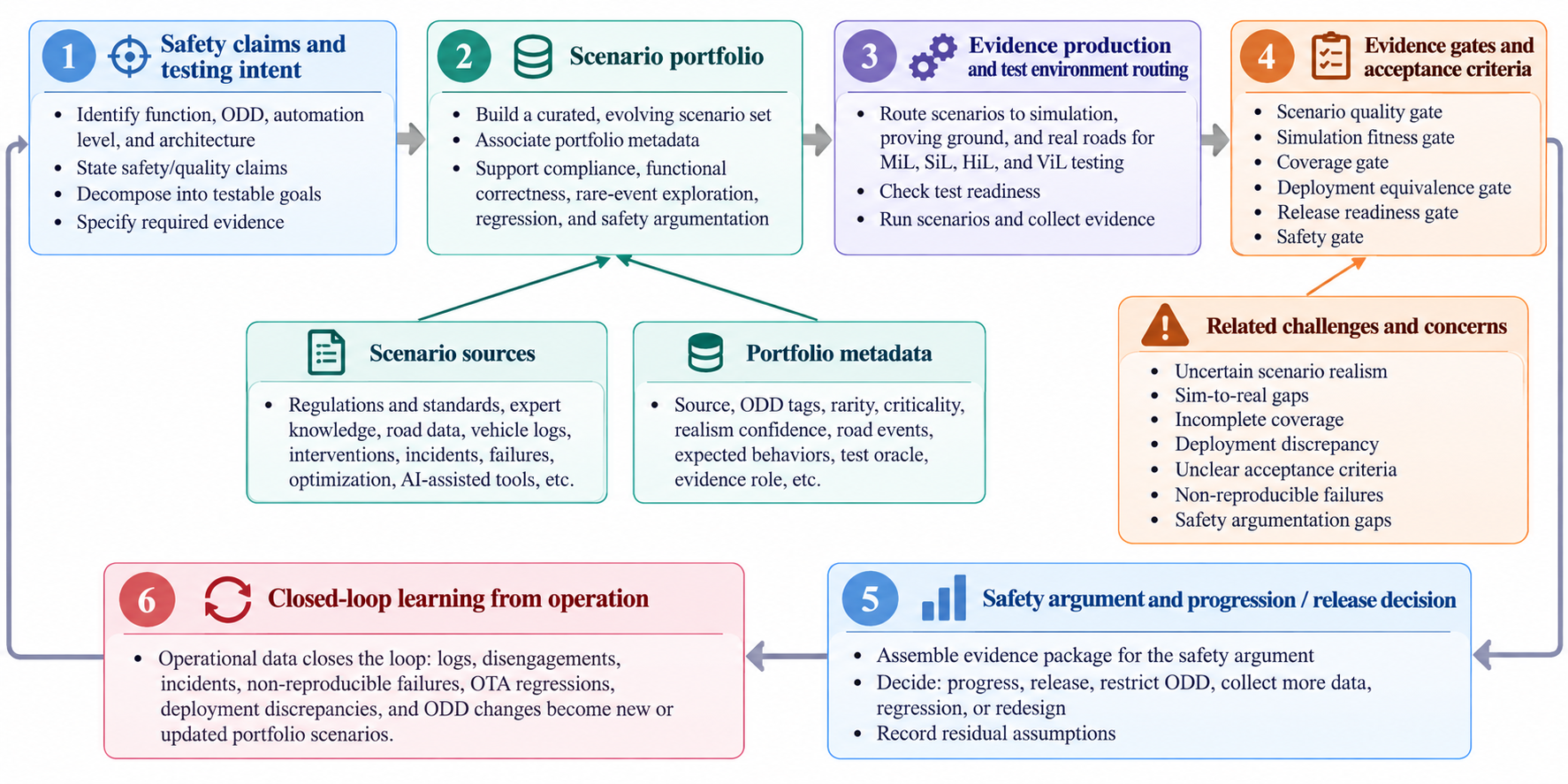}
    \caption{A practice-informed evidence-centered closed-loop ADS testing framework synthesized from the interview findings.}
    \label{fig:framework}
\end{figure*}

Building on the findings from RQ1 and RQ2, we synthesize a practice-informed framework that connects the main activities and limitations observed across the interviews. The framework is not intended to prescribe a universal testing process. Rather, it organizes the empirical findings into an evidence-centered structure that can guide how testing intent, scenarios, test environments, acceptance decisions, and operational feedback are connected.

Table~\ref{tab:framework-traceability} summarizes the empirical basis of each stage. Stages~1--3 primarily synthesize the practices identified in RQ1, while Stages~4--5 reflect how practitioners assess testing sufficiency together with the assurance limitations identified in RQ2. Stage~6 captures the recurring use of operational data and feedback to extend scenarios and regression testing. The six-stage organization itself, and the consolidation of individual practices into specific evidence gates, are our synthesis of these findings rather than a process reported verbatim by any single participant.

\begin{table}[tbp]
\centering
\caption{Traceability between the framework and interview findings.}
\label{tab:framework-traceability}
\begin{tabular}{p{0.16\columnwidth} p{0.72\columnwidth}}
\hline
\textbf{Stage} & \textbf{Empirical basis} \\
\hline
1 & Test planning based on functions, requirements, ODD, data, and testing objectives (RQ1-A) \\
2 & Scenario-based testing; scenario coverage, realism, and database limitations (RQ1-B, RQ2-A) \\
3 & Complementary MiL/SiL/HiL/ViL, simulation, proving-ground, and road testing; test readiness and execution constraints (RQ1-B, RQ2-B/C) \\
4 & Progression criteria, metrics, coverage, simulation fidelity, data quality, and deployment discrepancies (RQ1-B, RQ2-B/D) \\
5 & Acceptance criteria, safety argumentation, benchmarks, and progression/release decisions (RQ1-C, RQ2-D) \\
6 & Operational logs, interventions, failures, and newly encountered scenarios feeding subsequent testing (RQ1-A/B, RQ2-C) \\
\hline
\end{tabular}
\end{table}

\subsection{Stage 1: Define Safety Claims and Testing Intent}

Testing begins by clarifying the safety or quality objective to be supported and the evidence needed for it. Consistent with practitioners' emphasis on functions, requirements, ODD, and testing objectives, these factors should guide the definition of concrete testing goals and corresponding scenarios, environments, measures, and acceptance criteria. This makes testing evidence-driven rather than starting from an arbitrary collection of test cases.

\subsection{Stage 2: Construct and Maintain a Scenario Portfolio}

Scenarios can originate from standards and regulations, expert knowledge, real-world data, optimization, and AI-assisted generation. Reflecting the reported challenges of coverage and realism, the portfolio should be continuously enriched and, where possible, associated with information such as source, ODD, criticality, realism, expected behavior, and testing purpose. This helps relate individual scenarios to the evidence they are intended to provide.

\subsection{Stage 3: Plan and Produce Evidence Across Test Environments}

Scenarios should be assigned to testing environments according to the evidence required and the trade-offs among realism, scalability, cost, efficiency, and safety. This reflects practitioners' use of complementary MiL, SiL, HiL, ViL, simulation, proving-ground, and public-road testing rather than a universally fixed sequence. Test readiness, including system configuration, logging, synchronization, data access, and ground truth, should also be established before execution.

\subsection{Stage 4: Apply Evidence Gates and Acceptance Criteria}

The produced evidence should be evaluated before progression. Based on the reported practices and challenges, relevant checks may concern scenario quality, simulation fitness, coverage, safety and performance, data quality, regression needs, deployment equivalence, and release readiness. These gates consolidate several criteria reported separately by participants; they are intended as a structured way of exposing where additional evidence may still be required.

\subsection{Stage 5: Build the Safety Argument and Decide Progression or Release}

Evidence from testing should then be assembled to support the intended safety and quality claims. Depending on its sufficiency, the system may progress, be released, operate under a restricted ODD, require additional testing or data collection, or undergo redesign. This stage reflects the finding that progression and acceptance depend on a combination of requirements, quantitative evidence, benchmarks, safety considerations, and engineering judgement rather than a universal threshold.

\subsection{Stage 6: Close the Loop with Operational Feedback}

Finally, evidence from road testing and deployment should feed back into subsequent testing. Vehicle logs, interventions, disengagements, failures, deployment discrepancies, and newly encountered situations can be used to extend the scenario portfolio and regression tests, enabling testing to evolve with the ADS throughout its lifecycle.

\subsection{Scope and Validation}

The framework is an analytical synthesis of practices and limitations reported by the interviewees; it has not yet been independently evaluated as an end-to-end process in an industrial project. We therefore do not claim that following the framework necessarily improves testing effectiveness or safety. Its current contribution is to make the relationships among testing intent, evidence production, evidence sufficiency, and operational feedback explicit and traceable to the empirical findings. Future work should evaluate its applicability, completeness, and effectiveness through industrial case studies or practitioner-based validation.

\section{Discussion}
\label{sec:discussion}

\subsection{Future Directions for ADS Testing}
\label{sec:discussion:future_trends}

The interview findings suggest three broader directions in which ADS testing may evolve. First, testing is likely to become increasingly \textit{automated and data-driven}. Participants anticipated greater use of operational data, automated test execution, AI-assisted scenario generation, and tighter feedback loops between deployed vehicles and development environments. P9, for example, envisioned continuous use of vehicle logs to feed subsequent testing and development, facilitated by software-defined vehicles and Over-the-Air (OTA) updates. P6 similarly anticipated more autonomous testing tools that could select tests, collect data, and identify problems with less manual intervention. Together, these developments may shorten testing cycles and allow test portfolios to evolve more continuously with observed system behavior.

Second, participants expected stronger integration of \textit{end-to-end and world-model-based approaches}. P7 pointed to increasing investment in world models capable of learning and generating driving environments and sensor interactions, potentially integrating simulation, training, and testing more closely. Participants also discussed generative AI and end-to-end approaches for scenario and sensor data generation. However, these technologies do not remove the evidence-quality problem identified in RQ2: their usefulness for testing still depends on fidelity, controllability, reliability, and, particularly for complex learned models, interpretability.

Third, ADS testing may move toward more \textit{transparent and systematic assurance}. Participants highlighted clearer performance measures, greater sharing of benchmarks and testing knowledge, and stronger pre-deployment safety justification. P3 and P8 expected greater transparency as ADS deployment becomes more widespread, while P1 anticipated reduced reliance on proven-in-use arguments as more systematic forms of scenario-based validation mature. These developments would shift attention from demonstrating that large amounts of testing were performed toward explaining why the resulting evidence is sufficient for the intended safety claims.

\subsection{Implications for Practitioners and Researchers}
\label{sec:discussion:implications}

For practitioners, the findings suggest that ADS testing should be organized around the evidence required for engineering and safety decisions rather than around a fixed sequence of test environments. In particular, the purpose and limitations of evidence produced in each environment should be made explicit: simulation may provide scalable coverage, HiL may expose hardware and communication effects, and vehicle testing may reveal integrated behavior and environmental effects that cannot be represented sufficiently elsewhere. Maintaining traceability among testing intent, scenarios, environments, evaluation criteria, and progression decisions can make evidence gaps visible earlier and help determine whether additional testing, a different environment, or an ODD restriction is warranted.

For researchers, the findings highlight problems that extend beyond generating additional test scenarios or improving individual testing techniques. Practitioners still need methods for assessing scenario sufficiency and realism, determining when a simulation environment is fit for a particular testing claim, evaluating deployment equivalence, and defining defensible acceptance criteria for learning-based ADS. Emerging approaches such as generative AI, end-to-end models, and world models should therefore be evaluated not only by their ability to produce scenarios or realistic data, but also by whether the resulting evidence is controllable, reproducible, interpretable, and useful for practical assurance and release decisions.

\subsection{Threats to Validity}
\label{sec:discussion:ttv}

We considered threats to validity~\cite{verdecchia2023threats, lago2024threats} throughout the study design and execution. Given the empirical and qualitative nature of the study, \textit{construct validity}~\cite{sjoberg2022construct} was a particular concern, as the interview questions and resulting interpretations needed to accurately reflect practitioners' experiences of ADS testing. To address this, we recruited participants with direct experience in ADS testing and provided them with the study goals and scope before the interviews. During the interviews, we clarified questions where needed and encouraged participants to provide explanations and examples. Participants were also invited to review and revise their transcripts, and we followed up when statements were missing or unclear.

To strengthen the credibility of the analysis, we used an inductive thematic analysis process in which codes were iteratively merged, refined, and organized into higher-level themes. The evidence-centered interpretation presented in this paper is a further synthesis of these themes rather than a structure explicitly reported by individual participants.

A further limitation concerns participant selection and the self-reported nature of the data. Seven of the nine participants were recruited through convenience sampling, which may bias the sample toward organizations and practitioners accessible to the research team. Moreover, the findings reflect participants' descriptions of their experiences rather than direct observation of company testing processes or independent inspection of testing artifacts. We mitigated these risks by recruiting additional participants purposively, seeking participants with different roles and experience levels, requesting concrete examples during the interviews, and allowing participants to review their transcripts. Nevertheless, some organizational practices or viewpoints may be underrepresented.

The main threat to \textit{external validity}~\cite{siegmund2015views} is the relatively small sample of nine participants. We mitigated this by recruiting participants from nine companies across six countries, covering different company sizes, roles, and levels of experience, and by using multiple recruitment strategies. During later interviews, many observations became increasingly repetitive and fewer new themes emerged, suggesting reasonable saturation within the scope of the study. Nevertheless, the findings should not be interpreted as representative of the entire ADS industry, and practices may vary across organizations, regions, system architectures, and application domains.

\section{Conclusion}
\label{sec:conclusion}

As ADS continue to advance, testing must provide broad coverage of system behavior and credible evidence to support engineering and release decisions. Through interviews with practitioners from nine companies across six countries, we examined how ADS testing is organized in industrial practice and where confidence in the resulting evidence is limited.

Our findings characterize ADS testing as an evidence-building process in which practitioners determine what evidence is needed, produce and evaluate it across complementary testing environments, and decide whether it is sufficient for progression or acceptance. This process is constrained by recurring limitations related to scenario adequacy, simulation and deployment fidelity, test execution and observability, and assurance and decision sufficiency. Although practitioners reported several approaches to address these limitations, important gaps remain, particularly around determining whether testing evidence is sufficiently comprehensive and convincing.

Building on these findings, we proposed a practice informed evidence-centered closed-loop testing framework that connects testing intent, scenario selection, evidence production, evaluation, decision-making, and operational feedback. Rather than prescribing a universal process, the framework provides a structured way to reason about the evidence needed, how to produce it, and where further evidence may be required. Overall, the study provides an industry-grounded view of ADS testing and highlights important directions for improving both testing practice and research on ADS assurance.

\section*{Ethical Considerations}
The study was reviewed and approved by the UCL Research Ethics Committee (Project ID 2671). Participants were informed about the study and participated voluntarily. All participants provided informed consent before the interviews (available on Zenodo~\cite{song_2026_21408632}). Their responses were anonymized and reported without identifying individual participants.

\section*{Use of Generative AI}
ChatGPT (OpenAI) was used to assist with language refinement, text organization, and figure creation (Figure~\ref{fig:methodology}--\ref{fig:framework}). All generated text was reviewed and revised by the authors, who remain fully responsible for the study design, data collection and analysis, interpretation of results, and conclusions.

\bibliographystyle{IEEEtran}
\bibliography{ref}

\end{document}